\documentclass[preprint,3p]{elsarticle}

\usepackage{amsmath,amsfonts}
\usepackage{natbib,hyperref}

\newcommand{\ud}{\mathrm{d}}
\newcommand{\ve}{\varepsilon}

\biboptions{sort&compress}

\journal{Physics Letters B}

\begin{document}

\begin{frontmatter}

\title{Twister quintessence scenario}

\author[kul]{Orest Hrycyna} 
\ead{hrycyna@kul.lublin.pl}
\author[oa,kac,pavia]{Marek Szyd{\l}owski}
\ead{uoszydlo@cyf-kr.edu.pl}

\address[kul]{Department of Theoretical Physics, The John Paul II Catholic
University of Lublin, \\ Al. Rac{\l}awickie 14,
20-950 Lublin, Poland}
\address[oa]{Astronomical Observatory, Jagiellonian University,
Orla 171, 30-244 Krak{\'o}w, Poland}
\address[kac]{Mark Kac Complex Systems Research Centre, Jagiellonian
University, Reymonta 4, 30-059 Krak{\'o}w, Poland}
\address[pavia]{Dipartimento di Fisica Nucleare e Teorica, Universit{\`a} degli
studi di Pavia, via A. Bassi 6, I-27100 Pavia, Italy}

\begin{abstract}
We study generic solutions in a non-minimally coupled to gravity scalar field
cosmology. It is shown that dynamics for both canonical and phantoms scalar
fields with the potential can be reduced to the dynamical system from which the
exact forms for an equation of the state parameter can be derived. We have found
the stationary solutions of the system and discussed their stability. Within
the large class of admissible solutions we have found a
non-degenerate critical points and we pointed out multiple attractor type of
trajectory travelling in neighborhood of three critical points at which we have
the radiation dominating universe, the barotropic matter dominating state and
finally the de Sitter attractor. We have demonstrated the stability of this
trajectory which we call the twister solution. Discovered evolutional path is
only realized if there exist the non-minimal coupling constant. We have found
simple duality relations between twister solutions in phantom and canonical
scalar fields in the radiation domination phase. For the twister trajectory we
have found an oscillating regime of approaching the de Sitter attractor.
\end{abstract}


%\maketitle

\end{frontmatter}

Observational evidences \cite{Knop:2003iy, Riess:2004nr} indicated that the
current universe is dominated by an exotic form of matter with negative pressure
called dark energy. The properties of dark energy component of the universe can
be characterized by the equation of the state coefficient
$w_{\phi}=p_{\phi}/\rho_{\phi}$, where $p_{\phi}$ and $\rho_{\phi}$ are pressure
and energy density of dark energy component, respectively. In modern cosmology
the scalar field plays the major role in modelling of dark energy and in
explanation of the accelerated expansion of the current universe (for review see
\cite{Copeland:2006wr}). In the simplest case the scalar field minimally coupled
to gravity is assumed (so-called the quintessence idea \cite{Ratra:1987rm,
Wetterich:1987fm, Caldwell:1997ii, Wang:1999fa}). In such a case the equation of
the state parameter is $w_{\phi}>-1$, but as it was noted by Caldwell
\cite{Caldwell:1999ew} the observational data also admit the possibility that
$w_{\phi}<-1$. Such a ``phantom'' form of a scalar field describing the dark
energy component has many peculiar properties such as, for example, a big-rip
singularity (energy density becomes infinite in finite time), the Lorentz
invariance condition is then violated etc. \cite{Caldwell:2003vq,
Dabrowski:2003jm}.

In this contribution we investigate dynamics of the cosmological model with the
scalar field non-minimally coupled to the gravity with the positive and negative
kinetic energy forms (i.e., canonical and phantom scalar fields) in the
background of the flat Friedmann-Robertson-Walker (FRW) geometry. We point out
interesting properties of a three phase model obtained within these class of
solutions. They are interesting because they are generic and interpolate three
physically important phases of the evolution of the universe, namely, radiation,
matter and dark energy domination in the evolution of the universe. Therefore,
this solutions can be treated as a natural extension of the quartessence idea
\cite{Kamenshchik:2001cp, Bilic:2001cg, Makler:2002jv, Bento:2002ps}. In
standard cosmology the expression ``radiation dominated universe'' implies a
universe dominated by photons. In this Letter, the meaning is different: it means
a universe with effective equation of state parameter $w_{eff}=1/3$ dominated by
non-minimally coupled scalar field. In this case the dynamics of the scale
factor mimics the evolution of the radiation dominated universe.

In our investigations we apply the dynamical systems methods in exploring
stationary states represented by critical points in the phase space as well as
their stability \cite{Hrycyna:2008gk, Szydlowski:2008in, Hrycyna:2007mq,
Hrycyna:2007gd, Faraoni:2006ik, Faraoni:2000gx, Belinsky:1985zd,
Barvinsky:1994hx, Barvinsky:1998rn, Barvinsky:2008ia, Setare:2008mb,
Setare:2008pc}. We characterize all generic scenarios appearing in the case of
the constant non-minimal coupling for both canonical and phantom scalar fields. 
In our dynamical study we relax the choice of the potential function. The
presented approach to study the dynamics with the dynamical form of the equation
of the state parameter is a different form the most popular one mainly used in
the confrontation of the assumed model with dynamical dark energy with the
observational data \cite{Chevallier:2000qy, Linder:2004ng}. While the authors
who estimate parameters from the observational data postulate at the very
beginning the form of the parameterization of the equation of state parameter
$w(z)$ as a function of the redshift $z$, in the presented approach such a form
is directly derived from the closed dynamics of the FRW model filled by the
non-minimally coupled scalar field. Moreover, basing on the twister solution
one can derive approximated forms of the effective equation of the state
parameter $w(z)$ in three characteristic phases of the evolution of the
universe, namely during the radiation, the barotropic matter and the dark energy
domination.

In the model under consideration we assume the spatially flat FRW universe
filled with the non-minimally coupled scalar field and barotropic fluid with
the equation of the state coefficient $w_{m}$. The action assumes following form
\begin{equation}
S = \frac{1}{2}\int \ud^{4}x \sqrt{-g} \Bigg(\frac{1}{\kappa^{2}}R - \ve
\Big(g^{\mu\nu}\partial_{\mu}\phi\partial_{\nu}\phi + \xi R \phi^{2}\Big) -
2U(\phi) \Bigg) + S_{m},
\end{equation}
where $\kappa^{2}=8\pi G$, $\ve = +1,-1$ corresponds to canonical and phantom
scalar field, respectively, the metric signature is $(-,+,+,+)$,
$R=6\left(\frac{\ddot{a}}{a}+\frac{\dot{a}^{2}}{a^{2}}\right)$ is the Ricci
scalar, $a$ is
the scale factor and a dot denotes differentiation with respect to the
cosmological time and $U(\phi)$ is the scalar field potential function. $S_{m}$
is the action for the barotropic matter part.

The dynamical equation for the scalar field we can obtain from the variation
$\delta S/\delta \phi = 0$
\begin{equation}
\ddot{\phi} + 3 H \dot{\phi} + \xi R \phi + \ve U'(\phi) =0,
\end{equation}
and energy conservation condition from the variation $\delta S/\delta g^{\mu\nu}=0$
\begin{equation}
\mathcal{E}= \ve \frac{1}{2}\dot{\phi}^{2} + \ve3\xi H^{2}\phi^{2} + \ve3\xi H
(\phi^{2})\dot{} + U(\phi) + \rho_{m} - \frac{3}{\kappa^{2}}H^{2}.
\end{equation}
Then conservation conditions read
\begin{eqnarray}
\frac{3}{\kappa^{2}}H^{2} & = & \rho_{\phi} + \rho_{m}, \\
\dot{H} & = & -\frac{\kappa^{2}}{2}\Big[(\rho_{\phi}+p_{\phi}) +
\rho_{m}(1+w_{m})\Big]
\end{eqnarray}
where the energy density and the pressure of the scalar field are
\begin{eqnarray}
\rho_{\phi} & = & \ve\frac{1}{2}\dot{\phi}^{2}+U(\phi)+\ve3\xi H^{2}\phi^{2} +
\ve 3\xi H (\phi^{2})\dot{},\\
p_{\phi} & = & \ve\frac{1}{2}(1-4\xi)\dot{\phi}^{2} - U(\phi) + \ve\xi
H(\phi^{2})\dot{} - \ve2\xi(1-6\xi)\dot{H}\phi^{2} -
\ve3\xi(1-8\xi)H^{2}\phi^{2} + 2\xi\phi U'(\phi).
\end{eqnarray}

In what follows we introduce the energy phase space variables
\begin{equation}
x\equiv \frac{\kappa \dot{\phi}}{\sqrt{6}H}, \quad
y\equiv\frac{\kappa\sqrt{U(\phi)}}{\sqrt{3}H}, \quad
z\equiv\frac{\kappa}{\sqrt{6}}\phi,
\end{equation}
which are suggested by the conservation condition
\begin{equation}
\frac{\kappa^{2}}{3H^{2}}\rho_{\phi} + \frac{\kappa^{2}}{3H^{2}}\rho_{m} =
\Omega_{\phi} + \Omega_{m} = 1
\end{equation}
or in terms of the newly introduced variables
\begin{equation}
\Omega_{\phi} = y^{2} + \ve\Big[(1-6\xi)x^{2}+6\xi(x+z)^{2}\Big] = 1
-\Omega_{m}.
\end{equation}

The acceleration equation can be rewritten to the form
\begin{equation}
\dot{H} = -\frac{\kappa^{2}}{2}\Big(\rho_{\rm{eff}}+p_{\rm{eff}}\Big) =
-\frac{3}{2}H^{2}(1+w_{\rm{eff}})
\end{equation}
where the effective equation of the state parameter reads
\begin{eqnarray}
w_{\rm{eff}}=\frac{1}{1-\ve6\xi(1-6\xi)z^{2}}\Big[ -1 +
\ve(1-6\xi)(1-w_{m})x^{2} + \ve2\xi(1-3w_{m})(x+z)^{2} + \nonumber \\ 
 + (1+w_{m})(1-y^{2}) -\ve2\xi(1-6\xi)z^{2} - 2\xi\lambda y^{2} z\Big]
\label{eq:weff}
\end{eqnarray}
where $\lambda = -\frac{\sqrt{6}}{\kappa}\frac{1}{U(\phi)}\frac{\ud U(\phi)}
{\ud\phi}$.

The dynamical system of the model under considerations is in the form
\cite{Szydlowski:2008in}
\begin{eqnarray}
x' & = & -(x-\ve\frac{1}{2}\lambda y^{2})\Big[1-\ve6\xi(1-6\xi)z^{2}\Big] + 
\frac{3}{2}\left(x+6\xi z\right) 
\bigg[ -\frac{4}{3} - 2\xi \lambda y^{2} z \nonumber \\ & & 
+ \ve(1-6\xi)(1-w_{m})x^{2} 
+\ve2\xi(1-3w_{m})\left(x+z\right)^{2} + (1+w_{m})(1-y^{2}) \bigg],\\
y' & = & y\left(2-\frac{1}{2}\lambda x\right)
 \Big[1-\ve6\xi(1-6\xi)z^{2}\Big] 
 +\frac{3}{2} y \bigg[ -\frac{4}{3} - 2\xi \lambda y^{2} z \nonumber
 \\ & &
 + \ve(1-6\xi)(1-w_{m})x^{2} + 
 \ve2\xi(1-3w_{m})\left(x+z\right)^{2}
 + (1+w_{m})(1-y^{2})\bigg], \\
z' & = & x \Big[1-\ve6\xi(1-6\xi)z^{2}\Big], \\
\lambda' & = & -\lambda^{2}(\Gamma - 1)x\Big[1-\ve6\xi(1-6\xi)z^{2}\Big],
\end{eqnarray}
where $\Gamma = \frac{U(\phi),_{\phi\phi}U(\phi)}{U(\phi),_{\phi}^{2}}$ and a prime denotes
differentiation with respect to time $\tau$ defined as
\begin{equation}
\frac{\ud}{\ud \tau} = \Big[1-\ve6\xi(1-6\xi)z^{2}\Big] \frac{\ud}{\ud \ln{a}}.
\label{eq:time}
\end{equation}
If $\lambda$ is constant then we obtain the scaling potential $\exp{(\lambda
\phi)}$ and the basic system reduces to the $3$-dimensional autonomous dynamical
system in the case of the model with the barotropic matter. In the case without
the matter the dynamical system is a $2$-dimensional autonomous one.

In the rest of the Letter we will assume the following form of the function
$\Gamma(\lambda)$
\begin{equation}
\Gamma(\lambda)= 1 - \frac{\alpha}{\lambda^{2}}
\label{eq:18}
\end{equation}
where $\alpha$ is an arbitrary constant beside of the case $\alpha = 0$ 
for which $\Gamma =1$ which corresponds to an exponential potential. 
For assumed form of $\Gamma(\lambda)$ function we can simply eliminate
one of the variables namely $z$ given by the relation
\begin{equation}
z(\lambda)= - \int\frac{\ud \lambda}{\lambda^{2}\big(\Gamma(\lambda)-1\big)} =
\frac{\lambda}{\alpha} + \rm{const}
\end{equation}
where in the rest of the Letter we take the integration constant as equal to
zero.

From Eq. (\ref{eq:18}) and the definition of the function $\Gamma$ we
can simply calculate the form of the potential function
\begin{equation}
U(\phi) = U_{0}
\exp{\left[-\frac{\kappa^{2}}{6}\left(\frac{\alpha}{2}\phi^{2}+\beta\phi\right)\right]}
=
\tilde{U}_{0}\exp{\left[-\alpha\frac{\kappa^{2}}{12}\left(\phi+\frac{\beta}{\alpha}\right)^{2}\right]}
\end{equation}
where $\beta$ is the integration constant. As we can see the dynamics of the
model does not depend on the value of this parameter. In such a case we are
exploring the solutions in the very rich family of potential functions.

Following the Hartman-Grobman theorem \cite{Perko:2001} the system can be well 
approximated by the linear part of the system around a non-degenerate critical 
point. Then stability of the critical point is determined by eigenvalues of a
linearization matrix only. In Table \ref{tab:1} we have gathered critical points
appearing in twister scenario together with the eigenvalues of the
linearization matrix calculated at those points.

\begin{table}
\caption{The location and eigenvalues of the critical points in twister
quintessence scenario}
\label{tab:1}
\begin{center}
\begin{tabular}{|c|c|c|c|c|}
\hline
 $w_{\rm{eff}}$ & $w_{\phi}$ & $\Omega_{\phi}$ & location & eigenvalues \\
\hline
 $\frac{1}{3}$ & $\frac{1}{3}$ & $1$ & $x_{1}^{*}=0, y_{1}^{*}=0,
(\lambda_{1}^{*})^{2}=\frac{\alpha^{2}}{\ve6\xi}$ & $l_{1}=-6\xi$,
$l_{2}=12\xi$, $l_{3}=6\xi(1-3w_{m})$ \\
 $w_{m}$ & --- & $0$ & $x_{2}^{*}=0, y_{2}^{*}=0, \lambda_{2}^{*}=0$ &
$l_{1,3}=-\frac{3}{4}(1-w_{m})\Big(1\pm\sqrt{1-\frac{16}{3}\xi\frac{1-3w_{m}}{(1-w_{m})^{2}}}\Big)$,
$l_{2}=\frac{3}{2}(1+w_{m})$\\
 $-1$ & $-1$ & $1$ &  $x_{3a}^{*}=0, (y_{3a}^{*})^{2}=1, \lambda_{3a}^{*}=0 $ &
$l_{1,3}=-\frac{1}{2}\Big(3\pm\sqrt{9+\ve2\alpha-48\xi}\Big)$, $l_{2}=-3(1+w_{m})$\\
 $-1$ & $-1$ & $1$ & $x_{3b}^{*}=0$,
 $(y_{3b}^{*})^{2}=\frac{1}{\alpha}\ve24\xi$, &
 $l_{1}=-18\xi(1+w_{m})\left(1+\frac{\ve}{\alpha}4(1-6\xi)\right)$ \\
 & & & $(\lambda_{3b}^{*})^{2}=\alpha\left(\frac{\alpha}{\ve6\xi}-4\right)$ & 
 $l_{2,3}=-3\xi\left(1+\frac{\ve}{\alpha}4(1-6\xi)\right) \left(3\pm
 \sqrt{\frac{-7+\frac{\ve}{\alpha}12(3+14\xi)}{1+\frac{\ve}{\alpha}4(1-6\xi)}}\right)$\\
\hline
\end{tabular}
\end{center}
\end{table}

The critical point of a saddle type which represents the radiation dominated
universe $w_{\rm{eff}}=1/3$ is
$(x_{1}^{*}=0, y_{1}^{*}=0, (\lambda_{1}^{*})^{2}=\frac{\alpha^{2}}{\ve6\xi})$ and the
linearized solutions in the vicinity of this critical point are
\begin{equation}
\begin{array}{ccl}
x_{1}(\tau) & = & \frac{1}{2-3w_{m}}\Bigg\{
\Big[x_{1}^{(i)}-\frac{1}{\alpha}(1-3w_{m})(\lambda_{1}^{(i)}-\lambda_{1}^{*})\Big]\exp{(l_{1}\tau)} +
(1-3w_{m})\Big[x_{1}^{(i)}+\frac{1}{\alpha}(\lambda_{1}^{(i)}-\lambda_{1}^{*})\Big]
\exp{(l_{3}\tau)}\Bigg\},\\
y_{1}(\tau) & = & y_{1}^{(i)} \exp{(l_{2}\tau)}, \\
\lambda_{1}(\tau) & = & \lambda_{1}^{*} - \frac{\alpha}{2-3w_{m}} \Bigg\{
\Big[x_{1}^{(i)}-\frac{1}{\alpha}(1-3w_{m})(\lambda_{1}^{(i)}-\lambda_{1}^{*})
\Big]\exp{(l_{1}\tau)} -
\Big[x_{1}^{(i)}+\frac{1}{\alpha}(\lambda_{1}^{(i)}-\lambda_{1}^{*})\Big]
\exp{(l_{3}\tau)} \Bigg\},
\end{array}
\end{equation}
where $l_{1}=-6\xi$, $l_{2}=12\xi$ and $l_{3}=6\xi(1-3w_{m})$ are eigenvalues of
the linearization matrix calculated at this critical point and the
transformation from time $\tau$ into the scale factor can be made using relation
(\ref{eq:time}) calculated at the critical point
$$
\ln{\left(\frac{a}{a^{(i)}_{1}}\right)} =
\int_{0}^{\tau}\Big[1-\ve6\xi(1-6\xi)z(\lambda^{*}_{1})^{2}\Big]\ud\tau =
6\xi\tau,
$$
where $a^{(i)}_{1}$ is the initial value of the scale factor at $\tau=0$.
In the case of canonical scalar field $\ve=+1$ this critical point
exists only if $\xi>0$ and for the phantom scalar field $\ve=-1$ if $\xi<0$.

The matter dominated universe where $w_{\rm{eff}}=w_{m}$ is represented by the
critical point
$(x_{2}^{*}=0,y_{2}^{*}=0,\lambda_{2}^{*}=0)$ which character depends on the value
of the parameter $d$
$$
d=1-\frac{16}{3}\xi\frac{1-3w_{m}}{(1-w_{m})^{2}}.
$$

For $d>0$ the critical point is of a saddle type and the linearized solutions
are in the form
\begin{equation}
\begin{array}{ccl}
x_{2}(\tau) & = & \frac{1}{2\sqrt{d}}\bigg\{ (1+\sqrt{d})\Big[
x_{2}^{(i)}+\frac{1}{\alpha}\frac{3}{4}(1-w_{m})(1-\sqrt{d})\lambda_{2}^{(i)} 
\Big] \exp{(l_{1}\tau)} - \\
& & \qquad \qquad  - 
(1-\sqrt{d})\Big[
x_{2}^{(i)}+\frac{1}{\alpha}\frac{3}{4}(1-w_{m})(1+\sqrt{d})\lambda_{2}^{(i)} 
\Big] \exp{(l_{3}\tau)} \bigg\}, \\
y_{2}(\tau) & = & y_{2}^{(i)} \exp{(l_{2}\tau)}, \\
\lambda_{2}(\tau) &= & -\frac{2\alpha}{3(1-w_{m})\sqrt{d}}
\bigg\{
\Big[
x_{2}^{(i)}+\frac{1}{\alpha}\frac{3}{4}(1-w_{m})(1-\sqrt{d})\lambda_{2}^{(i)} 
\Big]
\exp{(l_{1}\tau)} -  \\
& & \qquad \qquad \qquad \quad - 
\Big[
x_{2}^{(i)}+\frac{1}{\alpha}\frac{3}{4}(1-w_{m})(1+\sqrt{d})\lambda_{2}^{(i)} 
\Big] \exp{(l_{3}\tau)}
\bigg\}.
\end{array}
\end{equation}
where
$l_{1,3}=-\frac{3}{4}(1-w_{m})\Big(1\pm\sqrt{1-\frac{16}{3}\xi\frac{1-3w_{m}}{(1-w_{m})^{2}}}\Big)$,
$l_{2}=\frac{3}{2}(1+w_{m})$ are eigenvalues of the linearization matrix.

For $d<0$ the critical point is of an unstable focus type
\begin{equation}
\begin{array}{ccl}
x_{2}(\tau) & = & -\frac{\exp{\big(-\frac{3}{4}(1-w_{m})\tau\big)}}{\sqrt{|d|}} \bigg\{ 
\Big[x_{2}^{(i)} +
\frac{1}{\alpha}\frac{3}{4}(1-w_{m})(1+|d|)\lambda_{2}^{(i)}
\Big]\sin{\bigg(\frac{3}{4}(1-w_{m})\sqrt{|d|}\tau\bigg)} - \\
 & & \qquad \qquad \qquad \quad - \sqrt{|d|} x_{2}^{(i)}
 \cos{\bigg(\frac{3}{4}(1-w_{m})\sqrt{|d|}\tau \bigg)}
\bigg\} , \\
y_{2}(\tau) & = & y_{2}^{(i)} \exp{\big(\frac{3}{2}(1+w_{m})\tau\big)} , \\
\lambda_{2}(\tau) & = & 
\frac{4}{3}\frac{\alpha
\exp{\big(-\frac{3}{4}(1-w_{m})\tau\big)}}{(1-w_{m})\sqrt{|d|}} \bigg\{
\Big[x_{2}^{(i)} +
 \frac{1}{\alpha}\frac{3}{4}(1-w_{m})\lambda_{2}^{(i)}
 \Big]\sin{\bigg(\frac{3}{4}(1-w_{m})\sqrt{|d|}\tau\bigg)} + \\
  & & \qquad \qquad \qquad \quad  + 
  \frac{1}{\alpha}\frac{3}{4}(1-w_{m})\sqrt{|d|}\lambda_{2}^{(i)}
  \cos{\bigg(\frac{3}{4}(1-w_{m})\sqrt{|d|}\tau\bigg)} \bigg\}.
\end{array}
\end{equation}
For both cases the transformation from time $\tau$ to the scale factor in the
vicinity of the critical point corresponding to matter dominated universe is the
following
$$
\ln{\left(\frac{a}{a^{(i)}_{2}}\right)}=\tau,
$$
where $a^{(i)}_{2}$ is the initial value of the scale factor at $\tau=0$.

The final critical point represents the de Sitter universe with
$w_{\rm{eff}}=-1$ is $(x_{3a}^{*}=0,(y_{3a}^{*})^{2}=1,\lambda_{3a}^{*}=0)$ its
character depends on the value of the discriminant $\Delta_{3a}=9+\ve2\alpha-48\xi$ 
of the characteristic equation.

For $\Delta_{3a}<0$ the critical point is of a stable focus type and the
linearized solutions are
\begin{equation}
\begin{array}{ccl}
x_{3a}(\tau) & = &-\frac{\exp{(-\frac{3}{2}\tau)}}{\sqrt{|\Delta_{3a}|}}
\bigg\{
\Big[3x_{3}^{(i)}+\frac{9+|\Delta_{3a}|}{2\alpha}\lambda_{3}^{(i)}\Big]
\sin{\bigg(\frac{\sqrt{|\Delta_{3a}|}}{2}\tau\bigg)} 
- x_{3}^{(i)}\sqrt{|\Delta_{3a}|}\cos{\bigg(\frac{\sqrt{|\Delta_{3a}|}}{2}\tau\bigg)}
\bigg\}, \\
y_{3a}(\tau) & = & y_{3}^{*} +
(y_{3}^{(i)}-y_{3}^{*})\exp{\big(-3(1+w_{m})\tau\big)} , \\
\lambda_{3a}(\tau) & = &\frac{2\alpha
\exp{(-\frac{3}{2}\tau)}}{\sqrt{|\Delta_{3a}|}}
\bigg\{
\Big(x_{3}^{(i)}+\frac{3}{2\alpha}\lambda_{3}^{(i)}\Big)
\sin{\bigg(\frac{\sqrt{|\Delta_{3a}|}}{2}\tau\bigg)} + 
\frac{\sqrt{|\Delta|}}{2\alpha}\lambda_{3}^{(i)}
\cos{\bigg(\frac{\sqrt{|\Delta_{3a}|}}{2}\tau\bigg)}
\bigg\}.
\end{array}
\end{equation}

\begin{figure}
\begin{center}
\includegraphics[scale=0.75]{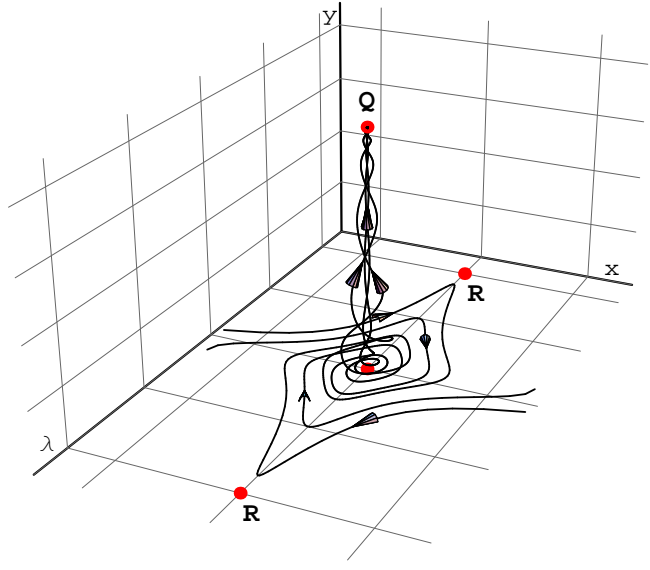}
\end{center}
\caption{The three-dimensional phase portrait of the dynamical system under
consideration for the canonical scalar field $\ve=+1$, the positive coupling
constant $\xi=6$ and $\alpha=-6$. Trajectories represent a twister type solution
which interpolates
between the radiation dominated universe $R$ (a saddle type critical point), the
matter
dominated universe (an unstable focus critical point) and the accelerating
universe $Q$ (a stable focus critical point).}
\label{fig:1}
\end{figure}

For $0<\Delta_{3a}<9$ the critical point is of a stable node type and when
$\Delta_{3a}>9$ is of a saddle type. The linearized solutions in the vicinity of
these types of critical points are
\begin{equation}
\begin{array}{ccl}
x_{3a}(\tau) & = & \frac{1}{2\sqrt{\Delta_{3a}}} 
\bigg\{
(3+\sqrt{\Delta_{3a}})
\Big[x_{3a}^{(i)} +
\frac{1}{2\alpha}(3-\sqrt{\Delta_{3a}})\lambda_{3a}^{(i)}\Big] \exp{(l_{1}\tau)}
- \\ & & \qquad \qquad (3-\sqrt{\Delta_{3a}})
\Big[x_{3a}^{(i)} +
\frac{1}{2\alpha}(3+\sqrt{\Delta_{3a}})\lambda_{3a}^{(i)}\Big] \exp{(l_{3}\tau)}
\bigg\}, \\
y_{3a}(\tau) & = & y_{3a}^{*} + (y_{3a}^{(i)}-y_{3a}^{*}) \exp{(l_{2}\tau)} , \\
\lambda_{3a}(\tau) & = & -\frac{\alpha}{\sqrt{\Delta_{3a}}}
\bigg\{
\Big[x_{3a}^{(i)}+\frac{1}{2\alpha}(3-\sqrt{\Delta_{3a}})\lambda_{3a}^{(i)}\Big]
 \exp{(l_{1}\tau)}- 
\Big[x_{3a}^{(i)}+\frac{1}{2\alpha}(3+\sqrt{\Delta_{3a}})\lambda_{3a}^{(i)}\Big]
 \exp{(l_{3}\tau)}
\bigg\}.
\end{array}
\end{equation}
where $l_{1,3}=-\frac{1}{2}\big(3\pm\sqrt{9+\ve2\alpha-48\xi}\big)$ and
$l_{2}=-3(1+w_{m})$ are eigenvalues of the linearization matrix and using
relation (\ref{eq:time}) we can write down transformation form time $\tau$ to
the scale factor $a$
$$
\ln\left(\frac{a}{a^{(i)}_{3a}}\right) = \tau,
$$
where $a^{(i)}_{3a}$ is the initial value of the scale factor at $\tau=0$. The
phase diagram of the system with the de Sitter state in the form of a stable
focus critical point is presented in Fig.~\ref{fig:1}.

\begin{figure}
\begin{center}
\includegraphics[scale=0.75]{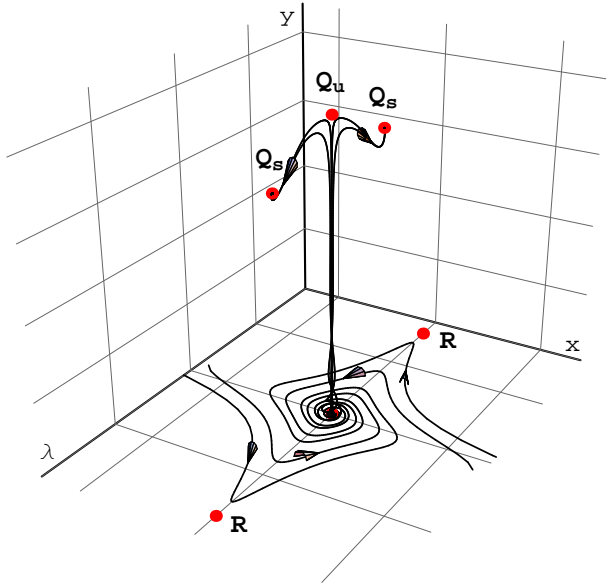}
\end{center}
\caption{The three-dimensional phase portrait of the investigated dynamical system
for the canonical scalar field $\ve=+1$, the coupling constant $\xi=6$ and
$\alpha=192$. Trajectories represent a twister typer solution interpolating
between the radiation dominated universe $R$ (a saddle type critical point), the
matter dominated universe (an unstable focus critical point), the accelerating
universe $Q_{u}$ represented by a saddle type critical point and final de Sitter
state $Q_{s}$ represented by a stable focus critical point.}
\label{fig:2}
\end{figure}

The next critical point represents also the de Sitter state with $w_{\rm{eff}}=-1$
($x^{*}_{3b}=0$, $(y^{*}_{3b})^{2}=\frac{1}{\alpha}\ve24\xi$,
$(\lambda^{*}_{3b})^{2}=\alpha\big(\frac{\alpha}{\ve6\xi}-4\big)$). The
linearized solution in the vicinity of this critical point can be presented in
the condensed form as
\begin{equation}
\left(\begin{array}{c}
x_{3b}(\tau)-x^{*}_{3b} \\ y_{3b}(\tau)-y^{*}_{3b} \\
\lambda_{3b}(\tau)-\lambda^{*}_{3b} \end{array}\right) = 
P_{3b} \left(\begin{array}{ccc}
\exp{(l_{1}\tau)} & 0 & 0 \\ 0 & \exp{(l_{2}\tau)} & 0 \\ 0 & 0 &
\exp{(l_{3}\tau)} 
\end{array}\right) P^{-1}_{3b} \left(\begin{array}{c} x^{(i)}_{3b}-x^{*}_{3b}
\\ y^{(i)}_{3b}-y^{*}_{3b} \\ \lambda^{(i)}_{3b}-\lambda^{*}_{3b}
\end{array}\right)
\end{equation}
where matrix 
\begin{equation}
P_{3b} = \left(\begin{array}{ccc}
-\frac{3}{\alpha}(1+w_{m}) &
-\frac{3}{2\alpha}-\frac{\sqrt{\Delta_{3b}}}{2\alpha} &
-\frac{3}{2\alpha}+\frac{\sqrt{\Delta_{3b}}}{2\alpha} \\
\frac{6(1+w_{m})\big(1+\frac{\ve}{\alpha}6(w_{m}-4\xi)\big)}{(1-3w_{m})y^{*}_{3b}\lambda^{*}_{3b}}
& \frac{y^{*}_{3b}\lambda^{*}_{3b}}{8\alpha}+\frac{\sqrt{\Delta_{3b}}}{8\alpha}
& \frac{y^{*}_{3b}\lambda^{*}_{3b}}{8\alpha}-\frac{\sqrt{\Delta_{3b}}}{8\alpha}
\\ 1 & 1 & 1 \end{array}\right),
\end{equation}
is constructed form the corresponding eigenvectors of the linearization matrix
of the system under considerations calculated at this critical point. The
eigenvalues are $l_{1}=-18\xi(1+w_{m})\big(1+\frac{\ve}{\alpha}4(1-6\xi)\big)$,
$l_{2,3}=-3\xi\big(1+\frac{\ve}{\alpha}4(1-6\xi)\big)\big(3\pm\sqrt{\Delta_{3b}}\big)$
and
$\Delta_{3b}=\frac{-7+\frac{\ve}{\alpha}12(3+14\xi)}{1+\frac{\ve}{\alpha}4(1-6\xi)}$.
The transformation from time $\tau$ to the scale factor $a$ from (\ref{eq:time})
is in the following form
$$
\ln{\left(\frac{a}{a^{(i)}_{3b}}\right)}=6\xi\Big(1+\frac{\ve}{\alpha}4(1-6\xi)\Big)\tau,
$$
where $a^{(i)}_{3b}$ is the initial value of the scale factor at $\tau=0$. The
phase space diagram for the system with one de Sitter state represented by a
saddle type critical point and two stable de Sitter states is presented in
Fig.~\ref{fig:2}. It is easy to check that if the critical point denoted as $3a$
is a saddle type, i.e. in the case when $\Delta_{3a}>9$, the critical point $3b$
is a stable one.

\begin{figure}
\begin{center}
\includegraphics[scale=0.75]{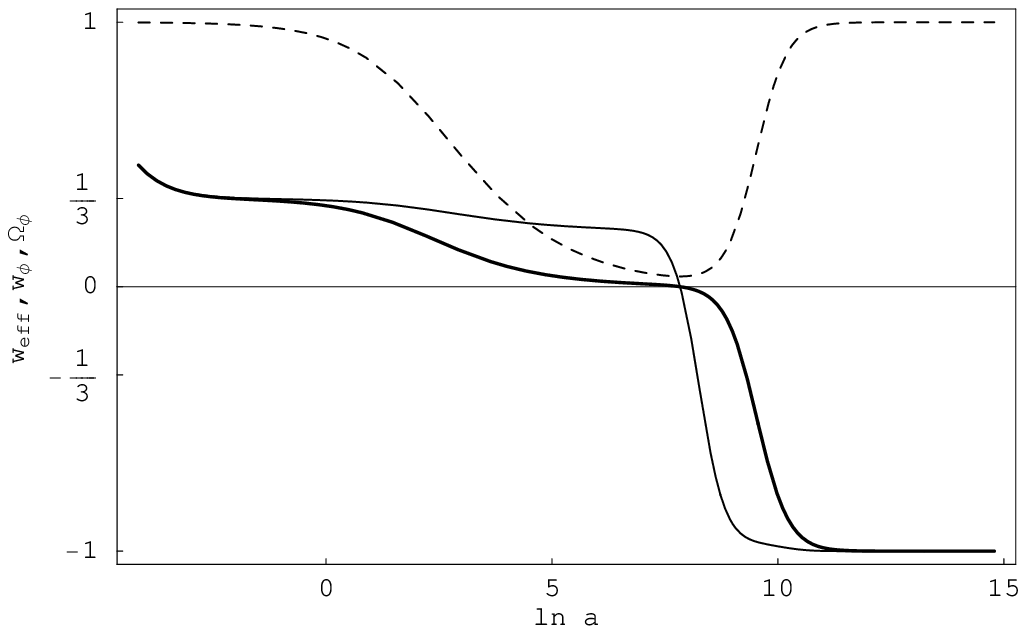}
\includegraphics[scale=0.75]{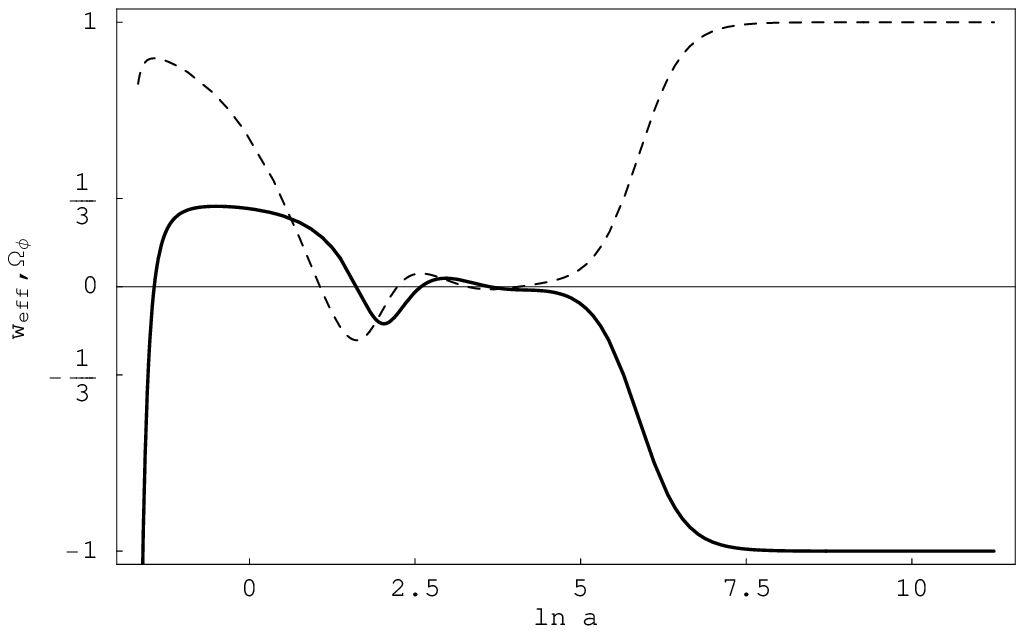}
\end{center}
\caption{The evolution of $w_{\rm{eff}}$ given by the relation
(\ref{eq:weff}) (thick line), $w_{\phi}$ -- thin line and $\Omega_{\phi}$
-- dashed line for the non-minimally
coupled canonical scalar field $\ve=+1$ and  $\xi=\frac{1}{8}$, $\alpha=-6$
(left) and $\xi=1$, $\alpha=-1$ (right). The existence of moments where 
$\Omega_{\phi}=0$ indicate singularities in $w_{\phi}$.
The sample
trajectories used to plot this relation start their evolution at $\ln{a}=0$
near the saddle type critical point ($w_{\rm{eff}}=1/3$) and then approach the
critical point representing the barotropic matter domination epoch
$w_{\rm{eff}}=w_{m}=0$ and next escape to the stable
de Sitter state with $w_{\rm{eff}}=-1$. The existence of a short
time interval during which $w_{\rm{eff}}\simeq\frac{1}{3}$ is the effect
of the nonzero coupling constant $\xi$ only.} 
\label{fig:3}
\end{figure}

The solutions of the linearized system in the vicinity of each critical point
$x_{i}(a)$, $y_{i}(a)$ and $\lambda_{i}(a)$ can be used to constrain the model
parameters through the cosmological data from various cosmological epochs. For
example, the parameters for the solution describing the radiation dominated
universe $(1)$ can be constrained from CMB data (its effects on CMB spectrum may
be different from pure photon gas \cite{Hu:2001bc}), and the solutions $(3)$
describing the current accelerating expansion of the universe through the SNIa
data. Therefore one can estimate the parameters of the variability with redshift
of true $w(a)$ (see Fig.~\ref{fig:3}). It is possible because we have the
linearization of the exact formula in different epochs.

The presented possibility of appearing the twister type
quintessence scenario is not restricted to the considered case of the
$\Gamma(\lambda)$ function (\ref{eq:18}). One can easily show that such a
scenario will be always possible if only the following functions calculated at
the critical points \cite{Szydlowski:2008in}
$$
f(\lambda^{*}) = (\lambda^{*})^{2}\big(\Gamma(\lambda^{*})-1\big) = {\rm const}, \qquad
\frac{\ud f(\lambda)}{\ud \lambda}\Big|_{\lambda^{*}} = f'(\lambda^{*}) =
{\rm const} 
$$
are finite.

In this Letter we pointed out the presence of the new interesting solution for 
the non-minimally coupled scalar field cosmology which we called the twister
solution (because of the shape of the corresponding trajectory in the phase
space, see Fig.~\ref{fig:1} and \ref{fig:2}).
This type of the solution is very interesting because in the phase space it
represents the 3-dimensional trajectory which interpolates different stages of 
evolution of the universe, namely, the radiation dominated, dust filled and
accelerating universe. We found linearized solutions around all these
intermediate phases and we are able to derive approximated forms of the
effective equation of the state parameter $w(a)$ in those epochs. It is
interesting that the presented structure of the phase space is allowed only for
a non-zero value of coupling constant, therefore it is a specific feature of the
non-minimally coupled scalar field cosmology.

\section*{Acknowledgments}
MS is very grateful to prof. Mauro Carfora for discussion and hospitality
during the visit in Pavia where this work was initiated.
This work has been supported by the Marie Curie Host Fellowships for the
Transfer of Knowledge project COCOS (Contract No. MTKD-CT-2004-517186).

\bibliography{twister}

\begin{thebibliography}{10}
\expandafter\ifx\csname url\endcsname\relax
  \def\url#1{\texttt{#1}}\fi
\expandafter\ifx\csname urlprefix\endcsname\relax\def\urlprefix{URL }\fi
\expandafter\ifx\csname href\endcsname\relax
  \def\href#1#2{#2} \def\path#1{#1}\fi

\bibitem{Knop:2003iy}
R.~A. Knop, G.~Aldering, R.~Amanullah, P.~Astier, G.~Blanc, M.~S. Burns,
  A.~Conley, S.~E. Deustua, M.~Doi, R.~Ellis, S.~Fabbro, G.~Folatelli, A.~S.
  Fruchter, G.~Garavini, S.~Garmond, K.~Garton, R.~Gibbons, G.~Goldhaber,
  A.~Goobar, D.~E. Groom, D.~Hardin, I.~Hook, D.~A. Howell, A.~G. Kim, B.~C.
  Lee, C.~Lidman, J.~Mendez, S.~Nobili, P.~E. Nugent, R.~Pain, N.~Panagia,
  C.~R. Pennypacker, S.~Perlmutter, R.~Quimby, J.~Raux, N.~Regnault,
  P.~Ruiz-Lapuente, G.~Sainton, B.~Schaefer, K.~Schahmaneche, E.~Smith, A.~L.
  Spadafora, V.~Stanishev, M.~Sullivan, N.~A. Walton, L.~Wang, W.~M.
  Wood-Vasey, N.~Yasuda, {New Constraints on $\Omega_M$, $\Omega_\Lambda$, and
  $w$ from an Independent Set of Eleven High-Redshift Supernovae Observed with
  HST}, Astrophys. J. 598 (2003) 102.
\newblock  \href{http://arxiv.org/abs/astro-ph/0309368}{{\tt
  arXiv:astro-ph/0309368}}.

\bibitem{Riess:2004nr}
A.~G. Riess, L.-G. Strolger, J.~Tonry, S.~Casertano, H.~C. Ferguson,
  B.~Mobasher, P.~Challis, A.~V. Filippenko, S.~Jha, W.~Li, R.~Chornock, R.~P.
  Kirshner, B.~Leibundgut, M.~Dickinson, M.~Livio, M.~Giavalisco, C.~C.
  Steidel, N.~Benitez, Z.~Tsvetanov, {Type Ia Supernova Discoveries at $z>1$
  From the Hubble Space Telescope: Evidence for Past Deceleration and
  Constraints on Dark Energy Evolution}, Astrophys. J. 607 (2004) 665--687.
\newblock  \href{http://arxiv.org/abs/astro-ph/0402512}{{\tt
  arXiv:astro-ph/0402512}}.

\bibitem{Copeland:2006wr}
E.~J. Copeland, M.~Sami, S.~Tsujikawa, {Dynamics of dark energy}, Int. J. Mod.
  Phys. D15 (2006) 1753--1936.
\newblock  \href{http://arxiv.org/abs/hep-th/0603057}{{\tt
  arXiv:hep-th/0603057}}.

\bibitem{Ratra:1987rm}
B.~Ratra, P.~J.~E. Peebles, {Cosmological consequences of a rolling homogeneous
  scalar field}, Phys. Rev. D37 (1988) 3406.

\bibitem{Wetterich:1987fm}
C.~Wetterich, {Cosmology and the fate of dilatation symmetry}, Nucl. Phys. B302
  (1988) 668.

\bibitem{Caldwell:1997ii}
R.~R. Caldwell, R.~Dave, P.~J. Steinhardt, {Cosmological Imprint of an Energy
  Component with General Equation of State}, Phys. Rev. Lett. 80 (1998)
  1582--1585.
\newblock  \href{http://arxiv.org/abs/astro-ph/9708069}{{\tt
  arXiv:astro-ph/9708069}}.

\bibitem{Wang:1999fa}
L.-M. Wang, R.~R. Caldwell, J.~P. Ostriker, P.~J. Steinhardt, {Cosmic
  Concordance and Quintessence}, Astrophys. J. 530 (2000) 17--35.
\newblock  \href{http://arxiv.org/abs/astro-ph/9901388}{{\tt
  arXiv:astro-ph/9901388}}.

\bibitem{Caldwell:1999ew}
R.~R. Caldwell, {A phantom menace? Cosmological consequences of a dark energy
  component with super-negative equation of state}, Phys. Lett. B545 (2002)
  23--29.
\newblock  \href{http://arxiv.org/abs/astro-ph/9908168}{{\tt
  arXiv:astro-ph/9908168}}.

\bibitem{Caldwell:2003vq}
R.~R. Caldwell, M.~Kamionkowski, N.~N. Weinberg, {Phantom Energy: Dark Energy
  with $w<-1$ Causes a Cosmic Doomsday}, Phys. Rev. Lett. 91 (2003) 071301.
\newblock  \href{http://arxiv.org/abs/astro-ph/0302506}{{\tt
  arXiv:astro-ph/0302506}}.

\bibitem{Dabrowski:2003jm}
M.~P. Dabrowski, T.~Stachowiak, M.~Szydlowski, {Phantom cosmologies}, Phys.
  Rev. D68 (2003) 103519.
\newblock  \href{http://arxiv.org/abs/hep-th/0307128}{{\tt
  arXiv:hep-th/0307128}}.

\bibitem{Kamenshchik:2001cp}
A.~Y. Kamenshchik, U.~Moschella, V.~Pasquier, {An alternative to quintessence},
  Phys. Lett. B511 (2001) 265--268.
\newblock  \href{http://arxiv.org/abs/gr-qc/0103004}{{\tt
  arXiv:gr-qc/0103004}}.

\bibitem{Bilic:2001cg}
N.~Bilic, G.~B. Tupper, R.~D. Viollier, {Unification of dark matter and dark
  energy: The inhomogeneous Chaplygin gas}, Phys. Lett. B535 (2002) 17--21.
\newblock  \href{http://arxiv.org/abs/astro-ph/0111325}{{\tt
  arXiv:astro-ph/0111325}}.

\bibitem{Makler:2002jv}
M.~Makler, S.~Quinet~de Oliveira, I.~Waga, {Constraints on the generalized
  Chaplygin gas from supernovae observations}, Phys. Lett. B555 (2003) 1.
\newblock  \href{http://arxiv.org/abs/astro-ph/0209486}{{\tt
  arXiv:astro-ph/0209486}}.

\bibitem{Bento:2002ps}
M.~C. Bento, O.~Bertolami, A.~A. Sen, {Generalized Chaplygin gas, accelerated
  expansion and dark energy-matter unification}, Phys. Rev. D66 (2002) 043507.
\newblock  \href{http://arxiv.org/abs/gr-qc/0202064}{{\tt
  arXiv:gr-qc/0202064}}.

\bibitem{Hrycyna:2008gk}
O.~Hrycyna, M.~Szydlowski, {Non-minimally coupled scalar field cosmology on the
  phase plane}, JCAP 04 (2009) 026.
\newblock  \href{http://arxiv.org/abs/0812.5096}{{\tt arXiv:0812.5096
  [hep-th]}}.

\bibitem{Szydlowski:2008in}
M.~Szydlowski, O.~Hrycyna, {Scalar field cosmology in the energy phase-space --
  unified description of dynamics}, JCAP 01 (2009) 039.
\newblock  \href{http://arxiv.org/abs/0811.1493}{{\tt arXiv:0811.1493
  [astro-ph]}}.

\bibitem{Hrycyna:2007mq}
O.~Hrycyna, M.~Szydlowski, {Route to Lambda in conformally coupled phantom
  cosmology}, Phys. Lett. B651 (2007) 8--14.
\newblock  \href{http://arxiv.org/abs/0704.1651}{{\tt arXiv:0704.1651
  [hep-th]}}.

\bibitem{Hrycyna:2007gd}
O.~Hrycyna, M.~Szydlowski, {Extended Quintessence with non-minimally coupled
  phantom scalar field}, Phys. Rev. D76 (2007) 123510.
\newblock  \href{http://arxiv.org/abs/0707.4471}{{\tt arXiv:0707.4471
  [hep-th]}}.

\bibitem{Faraoni:2006ik}
V.~Faraoni, M.~N. Jensen, {Extended quintessence, inflation, and stable de
  Sitter spaces}, Class. Quant. Grav. 23 (2006) 3005--3016.
\newblock  \href{http://arxiv.org/abs/gr-qc/0602097}{{\tt
  arXiv:gr-qc/0602097}}.

\bibitem{Faraoni:2000gx}
V.~Faraoni, {A crucial ingredient of inflation}, Int. J. Theor. Phys. 40 (2001)
  2259--2294.
\newblock  \href{http://arxiv.org/abs/hep-th/0009053}{{\tt
  arXiv:hep-th/0009053}}.

\bibitem{Belinsky:1985zd}
V.~A. Belinsky, I.~M. Khalatnikov, L.~P. Grishchuk, Y.~B. Zeldovich,
  {Inflationary stages in cosmological models with a scalar field}, Phys. Lett.
  B155 (1985) 232--236.

\bibitem{Barvinsky:1994hx}
A.~O. Barvinsky, A.~Y. Kamenshchik, {Quantum scale of inflation and particle
  physics of the early universe}, Phys. Lett. B332 (1994) 270--276.
\newblock  \href{http://arxiv.org/abs/gr-qc/9404062}{{\tt
  arXiv:gr-qc/9404062}}.

\bibitem{Barvinsky:1998rn}
A.~O. Barvinsky, A.~Y. Kamenshchik, {Effective equations of motion and initial
  conditions for inflation in quantum cosmology}, Nucl. Phys. B532 (1998)
  339--360.
\newblock  \href{http://arxiv.org/abs/hep-th/9803052}{{\tt
  arXiv:hep-th/9803052}}.

\bibitem{Barvinsky:2008ia}
A.~O. Barvinsky, A.~Y. Kamenshchik, A.~A. Starobinsky, {Inflation scenario via
  the Standard Model Higgs boson and LHC}, JCAP 11 (2008) 021.
\newblock  \href{http://arxiv.org/abs/0809.2104}{{\tt arXiv:0809.2104
  [hep-ph]}}.

\bibitem{Setare:2008mb}
M.~R. Setare, E.~N. Saridakis, {Braneworld models with a non-minimally coupled
  phantom bulk field: a simple way to obtain the -1-crossing at late times},
  JCAP 03 (2009) 002.
\newblock  \href{http://arxiv.org/abs/0811.4253}{{\tt arXiv:0811.4253
  [hep-th]}}.

\bibitem{Setare:2008pc}
M.~R. Setare, E.~N. Saridakis, {Non-minimally coupled canonical, phantom and
  quintom models of holographic dark energy}, Phys. Lett. B671 (2009) 331--338.
\newblock  \href{http://arxiv.org/abs/0810.0645}{{\tt arXiv:0810.0645
  [hep-th]}}.

\bibitem{Chevallier:2000qy}
M.~Chevallier, D.~Polarski, {Accelerating universes with scaling dark matter},
  Int. J. Mod. Phys. D10 (2001) 213--224.
\newblock  \href{http://arxiv.org/abs/gr-qc/0009008}{{\tt
  arXiv:gr-qc/0009008}}.

\bibitem{Linder:2004ng}
E.~V. Linder, Probing gravitation, dark energy, and acceleration, Phys. Rev.
  D70 (2004) 023511.
\newblock  \href{http://arxiv.org/abs/astro-ph/0402503}{{\tt
  arXiv:astro-ph/0402503}}.

\bibitem{Perko:2001}
L.~Perko, Differential Equations and Dynamical Systems, 3rd Edition, Texts in
  Applied Mathematics, Springer-Verlag, New York, 2001.

\bibitem{Hu:2001bc}
W.~Hu, S.~Dodelson, {Cosmic Microwave Background Anisotropies}, Ann. Rev.
  Astron. Astrophys. 40 (2002) 171--216.
\newblock  \href{http://arxiv.org/abs/astro-ph/0110414}{{\tt
  arXiv:astro-ph/0110414}}.

\end{thebibliography}
\bibliographystyle{elsarticle-num}

\end{document}